\documentclass[aps,prd,preprint,superscriptaddress,amsmath,amssymb,showpacs]{revtex4-1}
\usepackage{dcolumn}
\usepackage{graphicx}
\usepackage{float}
\usepackage{physics}
\usepackage[colorlinks=true,allcolors=blue]{hyperref}

\begin{document}

\title{Thermodynamics of heavy quarkonium in the spinning black hole background}

\author{Zhou-Run Zhu }
\email{zhuzhourun@zknu.edu.cn}
\affiliation{School of Physics and Telecommunications Engineering, Zhoukou Normal University, Zhoukou 466001, China}

\author{Sheng Wang}
\email{shengwang@mails.ccnu.edu.cn}
\affiliation{Institute of Particle Physics and Key Laboratory of Quark and Lepton Physics (MOS), Central China Normal University, Wuhan 430079, China}

\author{Xun Chen}
\email{chenxun@usc.edu.cn}
\affiliation{School of Nuclear Science and Technology, University of South China, Hengyang 421001, China}
\affiliation{Institute of Particle Physics and Key Laboratory of Quark and Lepton Physics (MOS), Central China Normal University, Wuhan 430079, China}

\author{Jun-Xia Chen}
\email{chenjunxia@mails.ccnu.edu.cn}
\affiliation{Institute of Particle Physics and Key Laboratory of Quark and Lepton Physics (MOS), Central China Normal University, Wuhan 430079, China}

\author{Defu Hou}
\email{houdf@mail.ccnu.edu.cn}
\affiliation{Institute of Particle Physics and Key Laboratory of Quark and Lepton Physics (MOS), Central China Normal University, Wuhan 430079, China}

\begin{abstract}
In this study, we explore the thermodynamics of heavy quarkonium within the context of a spinning black hole background. Specifically, we analyze the impact of angular momentum on various thermodynamic properties of heavy quarkonium, including interquark distance, free energy, binding energy, entropy, entropic force, and internal energy, based on the thermodynamic relationships. Our results demonstrate that angular momentum diminishes the maximum interquark distance, thereby facilitating quarkonium dissociation. Furthermore, we note that angular momentum suppresses free energy. The analysis of binding energy reveals that angular momentum enhances the dissociation of mesons into free quarks and antiquarks. Additionally, our findings indicate that angular momentum augments entropy and entropic force, thereby accelerating quarkonium dissociation. Angular momentum also increases internal energy at extended interquark distances. Lastly, we observe that the effects of angular momentum on quarkonium are more significant when the axis of the quark pair $Q\overline{Q}$ is perpendicular to the direction of angular momentum.
\end{abstract}
\maketitle

\section{Introduction}\label{sec:01_intro}

There is a lot of evidence suggesting that quark-gluon plasma (QGP) has been created during the heavy-ion collisions experiments \cite{Arsene:2004fa,Adcox:2004mh,Back:2004je,Adams:2005dq}. Heavy quarkonium is bounded by strong interaction and the dissociation of bound states implies the presence of hot and strongly coupled medium. The direct evidence is the suppression of heavy quarkonium because of Debye screening \cite{Matsui:1986dk}. Heavy quarkonium is considered one of the most sensitive probes for studying the properties of QGP.

AdS/CFT correspondence \cite{Maldacena:1997re,Witten:1998qj,Gubser:1998bc} is a valuable and non-perturbative tool that provides significant insights into studying the nature of strongly coupled matter. From the holographic perspective, heavy quarks are described by the endpoints of open strings on the boundary. The quark and antiquark pair is connected by the U-shape string which hangs down into the bulk. The single quark is dual to the straight string hanging from the boundary to the horizon. The dynamics of quark pairs or a single quark can be calculated from the Nambu-Goto action.

The free energy serves an important role in the theoretical research of color screening and heavy quarkonium. The authors of \cite{Rey:1998bq,Brandhuber:1998bs} first calculated the free energy of quark pair in the $\mathcal{N}$=4 SYM theory. The holographic investigations of free energy in the strongly coupled medium have been discussed in \cite{Andreev:2006eh,Boschi-Filho:2006hfm,Andreev:2006nw,Hayata:2012rw,Zhu:2023aaq}. The difference between the free energy of $Q\overline{Q}$ and two free quarks can be seen as the binding energy. The free energy and binding energy in non-conformal theories have been studied in \cite{Ewerz:2016zsx}. Moreover, the entropy is relevant to the nature of deconfinement \cite{Kharzeev:2014pha}. The results of lattice QCD \cite{Kaczmarek:2002mc,Petreczky:2004pz,Kaczmarek:2005zp} show that entropy has the maximum values near deconfinement temperature. Besides, the entropy increases with the interquark distance of $Q\overline{Q}$ and is constant at large separations. These observational results lead to the emergent entropic force, which is responsible for the dissociation of quarkonium. There are many attempts to study entropic force by using the holographic approach \cite{BitaghsirFadafan:2015zjc,Zhang:2016tem,Iatrakis:2015sua,Critelli:2016cvq,Dudal:2017max,Kioumarsipour:2021zyg,Zhang:2020zeo,Jena:2022nzw}.

In this paper, we aim to study the thermodynamics of heavy quarkonium in the spinning black hole background. Specifically, we will calculate the effect of angular momentum on the interquark distance, free energy, binding energy, entropy, entropic force, and internal energy of heavy quarkonium. Since the strongly coupled plasma produced in non-central heavy ion collisions may carry nonzero angular momentum, it is natural to consider the rotating effects on the physical quantities \cite{Liang:2004ph,Becattini:2007sr,Baznat:2013zx,STAR:2017ckg,Jiang:2016woz}. The confinement-deconfinement phase transition \cite{Chen:2020ath,Braga:2022yfe,Zhao:2022uxc}, energy loss \cite{Hou:2021own,Chen:2022obe,Chen:2023yug} and thermodynamics of heavy quarkonium \cite{Zhou:2021sdy,Wu:2022ufk} have been analysed in rotating background from holography. Other interesting works can be seen in \cite{McInnes:2014haa,McInnes:2016dwk,Cai:2023cjl,Zhang:2023psy,Zhao:2024ipr,Zhao:2023pne}.

Using the Lorentz transformation \cite{Erices:2017izj,BravoGaete:2017dso,Awad:2002cz}, the authors of \cite{Zhou:2021sdy,Wu:2022ufk} generalize the static background to a local rotating black hole and examine the thermodynamics of heavy quarkonium. However, the resulting black hole only denotes a small neighborhood around rotating radius $L$ and a domain range less than 2$\pi$ of the rotating matter. It is significant to investigate the thermodynamics of heavy quarkonium in the spinning black hole background \cite{Hawking:1998kw,Gibbons:2004ai,Gibbons:2004js,Garbiso:2020puw}. The Myers-Perry black holes are vacuum solutions under Einstein gravity and the authors of \cite{Garbiso:2020puw} study the hydrodynamic transport coefficients in these black holes background. As discussed in \cite{Garbiso:2020puw}, the boundary is compact and dual gauge theory locates on $S^3 \times \mathbb{R}$. Furthermore, it was noted that the large black holes (high temperature case) correspond to a spinning medium in flat space $\mathbb{R}^{3,1}$. If we consider the planar limit black brane, the dual gauge theory lives on a non-compact spacetime. Based on these, we will explore the thermodynamics of heavy quarkonium in the spinning black hole background.

The paper is organized as follows. In Sec.~\ref{sec:02}, we review the spinning Myers-Perry black hole background. In Sec.~\ref{sec:03}, we discuss the thermodynamics of heavy quarkonium in spinning black hole. In Sec.~\ref{sec:04}, we make the conclusion and discussion.

\section{Spinning Myers-Perry black hole background}\label{sec:02}

Before analyzing the impact of angular momentum on various thermodynamic properties of heavy quarkonium, we briefly review the work of Hawking et al on the spinning black hole background. The metric of the spinning black hole background is \cite{Hawking:1998kw}

\begin{equation}
\label{eqc1}
\begin{split}
ds^{2} & =-\frac{\Delta}{\rho^{2}}(dt_{H}-\frac{a\sin^{2}\theta_{H}}{\Xi_{a}}d\phi_{H}-\frac{b\cos^{2}\theta_{H}}{\Xi_{b}}d\psi_{H})^{2}\\
 & +\frac{\Delta_{\theta_{H}}\sin^{2}\theta_{H}}{\rho^{2}}(adt_{H}-\frac{r_{H}^{2}+a^{2}}{\Xi_{a}}d\phi_{H})^{2}+\frac{\Delta_{\theta_{H}}\cos^{2}\theta_{H}}{\rho^{2}}(bdt_{H}-\frac{r_{H}^{2}+b^{2}}{\Xi_{b}}d\psi_{H})^{2}+\frac{\rho^{2}}{\Delta}dr_{H}^{2}\\
 & -\frac{\rho^{2}}{\Delta_{\theta_{H}}}d\theta_{H}^{2}+\frac{1+\frac{r_{H}^{2}}{L^{2}}}{r_{H}^{2}\rho^{2}}(abdt_{H}-\frac{b\left(r^{2}+a^{2}\right)\sin^{2}\theta_{H}}{\Xi_{a}}d\phi_{H}-\frac{a\left(r^{2}+b^{2}\right)\cos^{2}\theta_{H}}{\Xi_{b}}d\psi_{H})^{2},
 \end{split}
\end{equation}
with
 \begin{equation}
\label{eqc11}
\begin{split}
\Delta=\frac{1}{r_{H}^{2}}(r_{H}^{2}+a^{2})(r_{H}^{2}+b^{2})(1+\frac{r_{H}^{2}}{L^{2}})-2M,\\
 \Delta_{\theta_{H}}=1-\frac{a^{2}}{L^{2}}\cos^{2}\theta_{H}-\frac{b^{2}}{L^{2}}\sin^{2}\theta_{H},\\
 \rho=r_{H}^{2}+a^{2}\cos^{2}\theta_{H}+b^{2}\sin^{2}\theta_{H},\\
 \Xi_{a}=1-\frac{a^{2}}{L^{2}},\\
\Xi_{b}=1-\frac{b^{2}}{L^{2}},\\
 \end{split}
\end{equation}
where $\phi_H$, $\psi_H$ and $\theta_H$ denote angular Hopf coordinates. $t_H$, $L$ and $r_H$ are time, AdS radius and AdS radial coordinate respectively. $a$ and $b$ denote angular momentum parameters. We consider $a=b$ in the calculations, namely spinning Myers-Perry black hole case \cite{Gibbons:2004ai,Gibbons:2004js}.

To simplify the calculation, one can use the more convenient coordinates and the mass are described by \cite{Murata:2008xr}
 \begin{equation}
\label{eqc111}
\begin{split}
t=t_{H},\\
 r^{2}=\frac{a^{2}+r_{H}^{2}}{1-\frac{a^{2}}{L^{2}}},\\
\theta=2\theta_{H},\\
 \phi=\phi_{H}-\psi_{H},\\
\psi=-\frac{2at_{H}}{L^{2}}+\phi_{H}+\psi_{H},\\
b=a,\\
\mu=\frac{M}{(L^{2}-a^{2})^{3}},\\
 \end{split}
\end{equation}

Then rewrite the expression of metric (\ref{eqc1}) as
 \begin{equation}
\label{eqc2}
\begin{split}
ds^{2}=-(1+\frac{r^{2}}{L^{2}})dt^{2}+\frac{dt^{2}}{G(\text{r)}}+\frac{r^{2}}{4}((\sigma^{1})^{2}+(\sigma^{2})^{2}+(\sigma^{3})^{2})+\frac{2\mu}{r^{2}}(dt+\frac{a}{2}\sigma^{3})^{2},
 \end{split}
\end{equation}
with
 \begin{equation}
\label{eqc3}
\begin{split}
G(r)=1+\frac{r^{2}}{L^{2}}-\frac{2\mu(1-\frac{a^{2}}{L^{2}})}{r^{2}}+\frac{2\mu a^{2}}{r^{4}},\\
\mu=\frac{r_{h}^{4}(L^{2}+r_{h}^{2})}{2L^{2}r_{h}^{2}-2a^{2}(L^{2}+r_{h}^{2})},\\
\sigma^{1}=-sin\psi dtd\theta+cos\psi sin\theta d\phi,\\
\sigma^{2}=cos\psi d\theta+sin\psi sin\theta d\phi,\\
\sigma^{3}=d\psi+cos\theta d\phi,\\
 \end{split}
\end{equation}
where
\begin{equation}
\label{eqc4}
\begin{split}
-\infty<t<\infty,\ r_{h}<r<\infty,\ 0\leq\theta\leq\pi,\ 0\leq\phi\leq2\pi,\ 0\leq\psi\leq4\pi.
 \end{split}
\end{equation}

One can use the coordinate transformation to obtain the planar black brane \cite{Garbiso:2020puw}
 \begin{equation}
\label{eqc5}
\begin{split}
t=\tau,\\
\frac{L}{2}(\phi-\pi)=x_{1},\\
\frac{L}{2}tan(\theta-\frac{\pi}{2})=x_{2},\\
\frac{L}{2}(\psi-2\pi)=x_{3},\\
r=\tilde{r},\\
 \end{split}
\end{equation}
where $(\tau, \widetilde{r}, x_1, x_2, x_3)$ represent new coordinates. Then scale these coordinates with a power of scaling factor $\beta$
 \begin{equation}
\label{eqc6}
\begin{split}
\tau\rightarrow\beta^{-1}\tau,\\
x_{1}\rightarrow\beta^{-1}x_{1},\\
x_{2}\rightarrow\beta^{-1}x_{2},\\
x_{3}\rightarrow\beta^{-1}x_{3},\\
\tilde{r}\rightarrow\beta\tilde{r},\\
\tilde{r_{h}}\rightarrow\beta\tilde{r_{h}}.(\beta\rightarrow \infty)\\
 \end{split}
\end{equation}

The metric of Schwarzschild black brane which is boosted in $\tau- x_3$ plane can be obtained as \cite{Garbiso:2020puw}
\begin{equation}
\label{eqc7}
\begin{split}
ds^{2}=\frac{r^{2}}{L^{2}}(-d\tau^{2}+dx_{1}^{2}+dx_{2}^{2}+dx_{3}^{2}+\frac{r_{h}^{4}}{r^{4}(1-\frac{a^{2}}{L^{2}})}(d\tau+\frac{a}{L}dx_{3})^{2})+\frac{L^{2}r^{2}}{r^{4}-r_{h}^{4}}dr^{2},
 \end{split}
\end{equation}
note that the Schwarzschild black brane can be recovered when $a=0$.

To simplify the calculation, we use the $z$ ($r=1/z$) as the holographic fifth coordinate
\begin{equation}
\label{eqc8}
\begin{split}
ds^{2}=\frac{1}{z^{2}L^{2}}[-d\tau^{2}+dx_{1}^{2}+dx_{2}^{2}+dx_{3}^{2}+\frac{z^{4}}{z_{h}^{4}(1-\frac{a^{2}}{L^{2}})}(d\tau+\frac{a}{L}dx_{3})^{2}]+\frac{L^{2}z_{h}^{4}}{z^{2}(z_{h}^{4}-z^{4})}dz^{2}.
 \end{split}
\end{equation}

The temperature is \cite{Garbiso:2020puw}
\begin{equation}
\label{eqb1}
T_h=\frac{\sqrt{L^{2}-a^{2}}}{z_{h}\pi L^{3}},
\end{equation}
where $z_h$ is the horizon.

\section{Thermodynamics of heavy quarkonium in spinning Myers-Perry black hole background}\label{sec:03}

In this section, we will study the thermodynamics of heavy quarkonium in spinning Myers-Perry black holes background.

The Nambu-Goto action is
\begin{equation}
\label{eq10}
S= -\frac{1}{2\pi\alpha'} \int d\tau d\sigma \sqrt{-det g_{\alpha \beta}},
\end{equation}
where $g_{\alpha \beta}$ is the determinant of induced metric.

It should be mentioned that the metric (\ref{eqc8}) is boosted in $\tau- x_3$ plane. Thus, we can define the direction of rotation is in $x_3$ axis. In the calculations, we consider the transverse case and parallel case respectively. The transverse case represents the axis of $Q\overline{Q}$ pair is on the $x_1 - x_2$ plane while parallel case denotes the axis of $Q\overline{Q}$ pair is on $x_3$ axis.

In transverse case, the world sheet coordinates are parameterized by
\begin{equation}
\label{eq11}
\tau=\xi,\ x_{1}=\eta,\ x_{2}=0,\ x_{3}=0,\ z=z(\eta),
\end{equation}

The Lagrangian density is
\begin{equation}
\label{eq12}
\mathcal{L}=\sqrt{A(z)+B(z)\dot{z}^{2}},
\end{equation}
with
 \begin{equation}
\label{eq13}
\begin{split}
A(z)=A(z_\perp)=\frac{1}{z^{4}L^{4}}-\frac{1}{L^{4}z_{h}^{4}(1-\frac{a^{2}}{L^{2}})},\
B(z)=B(z_\perp)=\frac{z_{h}^{4}}{z^{4}(z_{h}^{4}-z^{4})}-\frac{1}{(z_{h}^{4}-z^{4})(1-\frac{a^{2}}{L^{2}})}.
 \end{split}
\end{equation}

In parallel case, the world sheet coordinates are parameterized by
\begin{equation}
\label{eq111}
\tau=\xi,\ x_{1}=0,\ x_{2}=0,\ x_{3}=\eta,\ z=z(\eta).
\end{equation}

Then $A(z)$ and $B(z)$ are
 \begin{equation}
\label{eq13}
\begin{split}
A(z)=A(z_\parallel)=\frac{z_{h}^{4}-z^{4}}{z^{4}z_{h}^{4}L^{4}},\
B(z)=B(z_\parallel)=\frac{z_{h}^{4}}{z^{4}(z_{h}^{4}-z^{4})}-\frac{1}{(z_{h}^{4}-z^{4})(1-\frac{a^{2}}{L^{2}})}.
 \end{split}
\end{equation}

The interquark distance $L$ of particle pair is
 \begin{equation}
\label{eq14}
\begin{split}
L=2\int_{0}^{z_{c}}\sqrt{\frac{A(z_{c})B(z)}{A(z)^{2}-A(z)A(z_{c})}}dz,
 \end{split}
\end{equation}
where $z_c$ denotes the deepest point of the U-shaped string.

The free energy of $Q\overline{Q}$ is
 \begin{equation}
\label{eq15}
\begin{split}
\frac{\pi F_{Q\overline{Q}}}{\sqrt{\lambda}}=\int_{0}^{z_{c}}(\sqrt{\frac{A(z)B(z)}{A(z)-A(z_{c})}}-\sqrt{A(z \rightarrow0)})dz-\int_{z_{c}}^{\infty} \sqrt{A(z \rightarrow0)}dz,
 \end{split}
\end{equation}
where $\lambda$ is 't Hooft coupling and we set $\lambda=1$ in the calculations. In order to eliminate the divergence of free energy, one can consider the minimum subtraction \cite{Ewerz:2016zsx}.

The entropy of $Q\overline{Q}$ pair is
 \begin{equation}
\label{eq16}
\begin{split}
S=-\frac{\partial F_{Q\overline{Q}}}{\partial T}.
 \end{split}
\end{equation}

The free energy of a single quark is
 \begin{equation}
\label{eq17}
\begin{split}
\frac{F_{Q}}{\sqrt{\lambda}}=\frac{1}{2\pi}[\int_{0}^{z_{h}}(\sqrt{B(z)}-\frac{1}{z^{2}})dz-\frac{1}{z_{h}}].
 \end{split}
\end{equation}

As discussed in \cite{Ewerz:2016zsx}, the binding energy is $E_{Q\overline{Q}}=F_{Q\overline{Q}}-2F_{Q}$ and the internal energy is $U_{Q\overline{Q}}=F_{Q\overline{Q}}+TS_{Q\overline{Q}}$.

\begin{figure}[H]
    \centering
      \setlength{\abovecaptionskip}{0.1cm}
    \includegraphics[width=8.5cm]{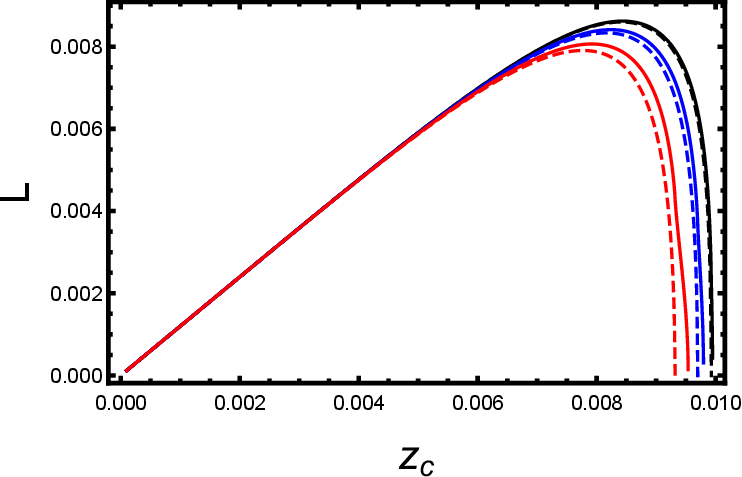}
    \caption{\label{fig1} Interquark distance $L$ of quarkonium versus $z_c$ with different angular momentum $a$. The black, blue and red line denote $a = 0.1,\ 0.2,\ 0.3$, respectively. The solid line (dashed line) represents the parallel (transverse) case.}
\end{figure}

In order to study the effect of angular momentum on the thermodynamics of heavy quarkonium, we perform calculations for various properties including interquark distance, free energy, binding energy, entropy, entropic force, and internal energy in the spinning black hole background. We take $L=1$ in the numerical calculations. As discussed in \cite{Garbiso:2020puw}, the background geometry presents a planar black brane in the large black hole (high temperature cases). We set a high temperature $T=100/\pi$. Moreover, we take $a=0.1, 0.2, 0.3$ as this spinning black brane exhibits stability when $a < 0.75L$.

In Fig.~\ref{fig1}, we depict the interquark distance $L$ of quarkonium versus $z_c$ with different angular momentum $a$. We observed that the interquark distance $L$ initially increases with $z_c$ until it reaches a maximum value $L_{m}$, which denotes the screening distance. This behavior indicates the characteristic U-shape string (connected string) feature. Subsequently, as we further increase $z_c$, the interquark distance decreases. This indicates that the U-shape string becomes unstable and starts to break into a straight string (disconnected string). Additionally, one can find that angular momentum reduces the maximum value of interquark distance $L_{m}$, implying that angular momentum promotes the dissociation of quarkonium. Importantly, it is found that the angular momentum has a stronger effect on quarkonium when the axis of $Q\overline{Q}$ is transverse to the direction of angular momentum.

In Ref. \cite{Zhou:2021sdy}, the authors investigate the impact of angular momentum on the interquark distance of quarkonium at T=0.1 GeV (low-temperature scenario). They observed that the quark-antiquark ($Q\overline{Q}$) pair is in the confinement phase at lower angular momentum, transitioning to the deconfinement phase at higher angular momentum. They noted that a maximum interquark distance emerges in the deconfinement phase (higher angular momentum scenario), whereas no such maximum exists in the confinement phase. Our result is not inconsistent with the result of Ref. \cite{Zhou:2021sdy}, which require sufficient angular momentum to reach the deconfinement phase. In our study, we concentrate on the effect of angular momentum on the interquark distance in a high-temperature scenario (deconfinement phase). Consequently, we find that the maximum interquark distance appears for any angular momentum values, and we observe that increasing angular momentum reduces this maximum interquark distance $L_{m}$.

\begin{figure}[H]
    \centering
      \setlength{\abovecaptionskip}{0.1cm}
    \includegraphics[width=15cm]{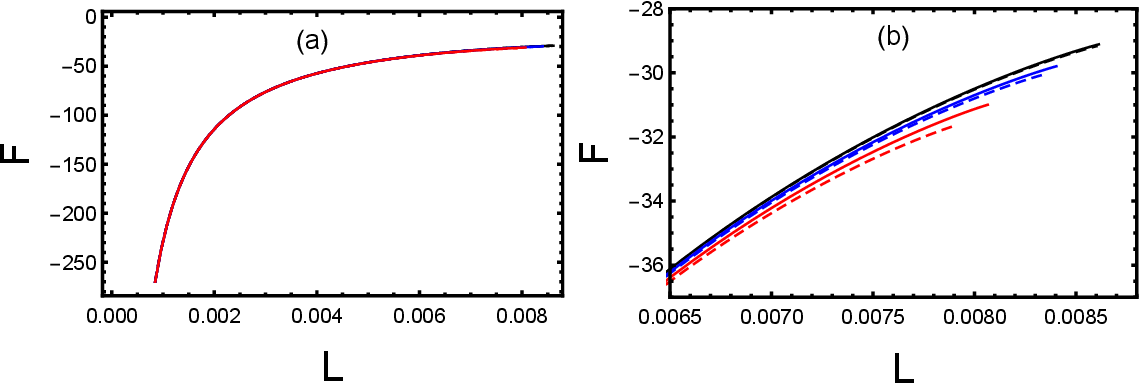}
    \caption{\label{fig2} Free energy $F$ of quarkonium versus $L$ with different angular momentum $a$. The black, blue and red line denote $a = 0.1,\ 0.2,\ 0.3$, respectively. The solid line (dashed line) represents the parallel (transverse) case. (b) is a partly enlarged view of (a).}
\end{figure}

In the following calculations, we will focus on the U-shape branch. In Fig.~\ref{fig2}, we plot the free energy $F$ of quarkonium versus $L$ with different angular momentum $a$. One can obviously observed that the free energy is negative in stable branch. This behaves as a Coulomb potential which can be analytically as $F\propto -1/L$. Furthermore, it is found that the angular momentum suppresses the free energy and has a larger effect in transverse case than in parallel.

\begin{figure}[H]
    \centering
      \setlength{\abovecaptionskip}{0.1cm}
    \includegraphics[width=15cm]{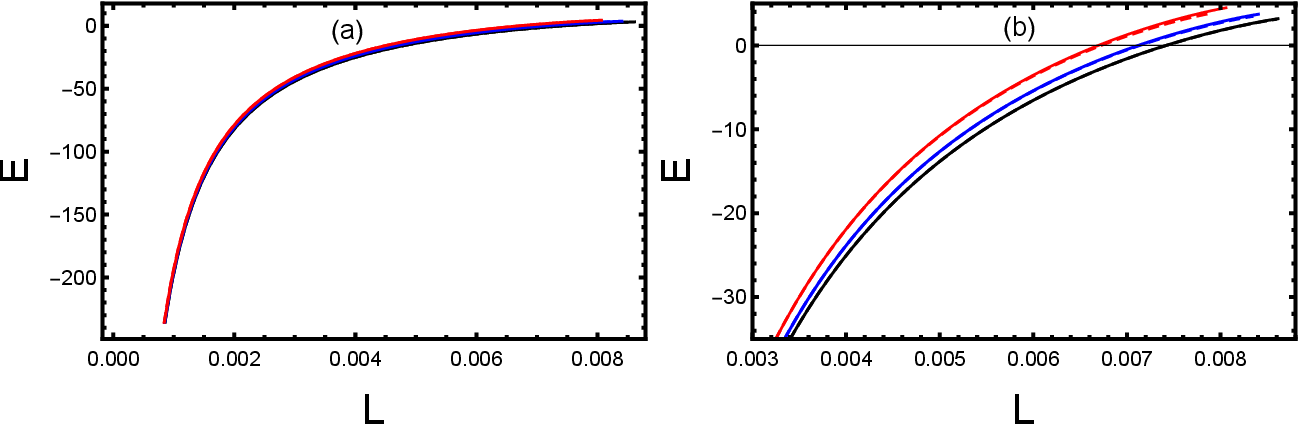}
    \caption{\label{fig3} Binding energy $E$ of quarkonium versus $L$ with different angular momentum $a$. The black, blue and red line denote $a = 0.1,\ 0.2,\ 0.3$, respectively. The solid line (dashed line) represents the parallel (transverse) case. (b) is a partly enlarged view of (a).}
\end{figure}

\begin{figure}[H]
    \centering
      \setlength{\abovecaptionskip}{0.1cm}
    \includegraphics[width=8.5cm]{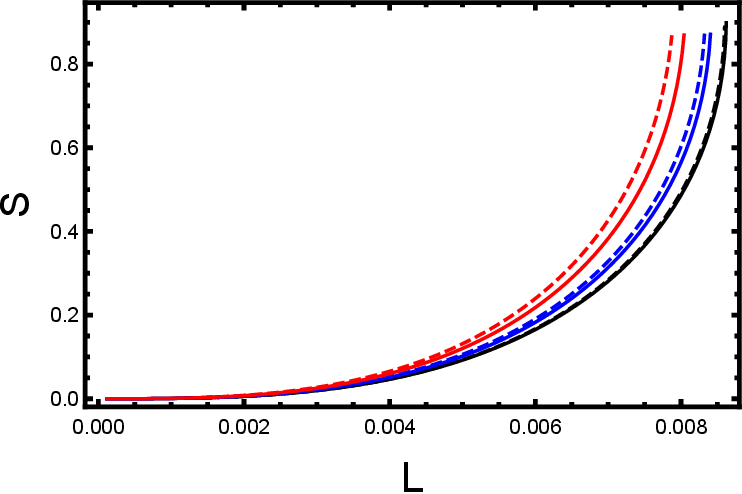}
    \caption{\label{fig4} Entropy $S$ of quarkonium versus $L$ with different angular momentum $a$. The black, blue and red line denote $a = 0.1,\ 0.2,\ 0.3$, respectively. The solid line (dashed line) represents the parallel (transverse) case. }
\end{figure}

In Fig.~\ref{fig3}, we plot binding energy $E$ of quarkonium versus $L$ with different angular momentum $a$. The binding energy represents the difference between the free energy of a quark pair and the free energy of single quarks. One can find that binding energy is less than zero at small $L$ which indicates $F_{Q\overline{Q}} < F_{Q}$. Binding energy approaches zero as increases $L$ and reaches $E(L_{c})=0$ at a critical value $L_{c}$ which suggests a phase transition between the connected string and the disconnected string. Binding energy is positive at large $L$ which means the free energy of a bound pair is larger than that of an unbound pair. Moreover, the angular momentum enhances the binding energy and reduces the critical value $L_{c}$. The physical implication is that the angular momentum favors the melting of meson into a free quark and antiquark.

\begin{figure}[H]
    \centering
      \setlength{\abovecaptionskip}{0.1cm}
    \includegraphics[width=8.5cm]{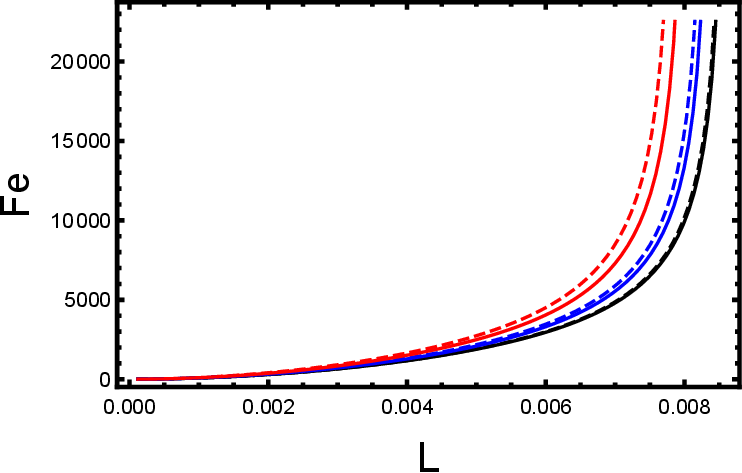}
    \caption{\label{fig5} Entropic force $F_e$ of quarkonium versus $L$ with different angular momentum $a$. The black, blue and red line denote $a = 0.1,\ 0.2,\ 0.3$, respectively. The solid line (dashed line) represents the parallel (transverse) case. }
\end{figure}

In Fig.~\ref{fig4}, we study the effect of angular momentum on entropy $S$. Fig.~\ref{fig4} shows that angular momentum leads to the increasing of the entropy. Besides, the entropy in the transverse case is larger than that of the parallel case. The entropic force is related to entropy through the equation $F_e = T \partial S/\partial L$. As is well-known, entropic force relies on the growth of the entropy which can be seen as a reason for the melting of quarkonium. Then we can infer that the angular momentum enhances the entropic force. The specific results are plotted in Fig.~\ref{fig5}. As expected, the angular momentum increases the entropic force and promotes the dissociation of quarkonium which is consistent with the results of \cite{Zhou:2021sdy,Wu:2022ufk}. One can also find that the entropic force is larger when the axis of $Q\overline{Q}$ is transverse to the direction of angular momentum than that in the parallel direction.

\begin{figure}[H]
    \centering
      \setlength{\abovecaptionskip}{0.1cm}
    \includegraphics[width=15cm]{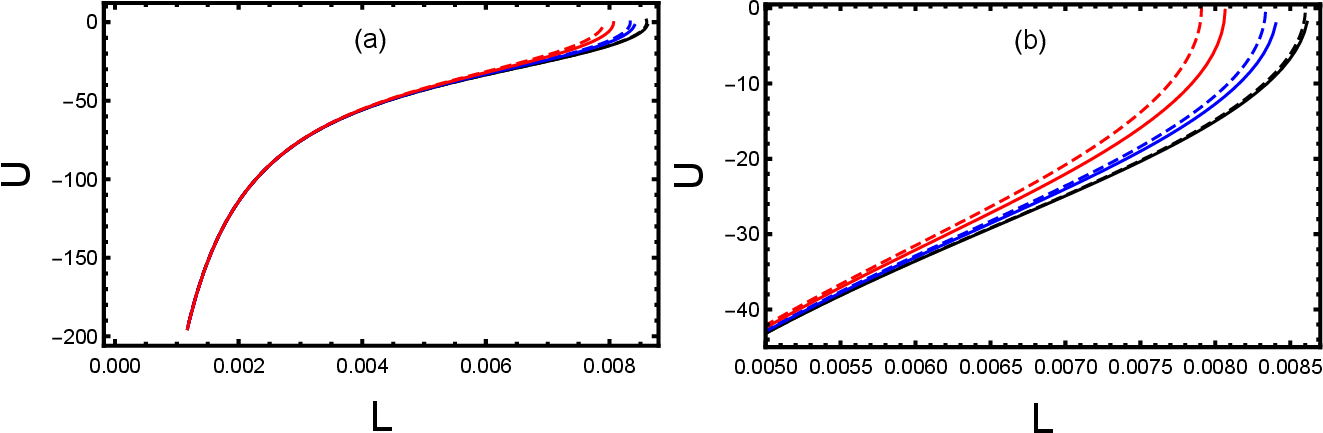}
    \caption{\label{fig6} Internal energy $U$ of quarkonium versus $L$ with different angular momentum $a$. The black, blue and red line denote $a = 0.1,\ 0.2,\ 0.3$, respectively. The solid line (dashed line) represents the parallel (transverse) case. (b) is a partly enlarged view of (a).}
\end{figure}

In Fig.~\ref{fig6}, we plot internal energy $U$ of quarkonium versus $L$ with different angular momentum $a$. The internal energy with different angular momentum is almost the same at small interquark distance $L$ while distinguishing at large $L$. The angular momentum increases the internal energy at large $L$ and has a larger effect in the transverse case than in the parallel case.

\section{Conclusion and discussion}\label{sec:04}

In this paper, we investigate the thermodynamics of heavy quarkonium within the context of a spinning black hole background. Specifically, we examine the impact of angular momentum on various thermodynamic properties of heavy quarkonium, including the interquark distance, free energy, binding energy, entropy, entropic force, and internal energy, through the lens of thermodynamic relationships.

The presence of angular momentum exerts diverse effects on quarkonium. Initially, it diminishes the maximum interquark distance $L_{m}$, thereby facilitating quarkonium dissociation. Analysis of the free energy reveals that angular momentum suppresses it. Conversely, angular momentum augments the binding energy and decreases the critical value $L_{c}$ , indicating a propensity for the dissociation of the meson into free quarks and antiquarks. Additionally, angular momentum escalates entropy and entropic force, thereby accelerating quarkonium dissociation. The internal energy remains largely unchanged at small interquark distances $L$, but diverges at larger $L$ with angular momentum increasing the internal energy at these distances.

In summary, angular momentum promotes the dissociation of quarkonium, aligning with findings from \cite{Zhou:2021sdy,Wu:2022ufk}. However, we observe that angular momentum exerts a more pronounced effect when the axis of  $Q\overline{Q}$ is transverse to the direction of angular momentum, contrary to the conclusions in \cite{Zhou:2021sdy}. Unlike \cite{Zhou:2021sdy}, which considers a local rotating black hole, our study delves into thermodynamics within a spinning black hole background. It is noteworthy that the spectral function and the influence of angular momentum on heavy quarkonium dissociation in a spinning black hole background have been previously discussed in \cite{Zhu:2024uwu}. According to \cite{Zhu:2024uwu}, angular momentum reduces the spectral function's peak height and broadens the peak width, signifying a tendency towards quarkonium dissociation. These findings corroborate our results.

Lastly, the investigation of configuration entropy in a spinning black hole background presents an intriguing avenue for future research. We intend to explore this topic in subsequent studies.

\section*{Acknowledgments}

Defu Hou's research is supported in part by the National Key Research and Development Program of China under Contract No. 2022YFA1604900. Additionally, he receives partial support from the National Natural Science Foundation of China (NSFC) under Grant No.12435009, and No. 12275104. Xun Chen is funded by the Natural Science Foundation of Hunan Province of China under Grant No. 2022JJ40344, as well as the Research Foundation of the Education Bureau of Hunan Province, China under Grant No. 21B0402. Xun Chen is also supported by the NSFC under Grant Nos. 12405154 and open Fund for Key Laboratories of the Ministry of Education under Grants Nos. QLPL2024P01. Zhou-Run Zhu's work is supported by the startup Foundation projects for Doctors at Zhoukou Normal University, with the project number ZKNUC2023018.


\section*{Appendix}

\subsection*{Appendix A}

In the appendix A, we demonstrate the first order phase transition temperature decreases with angular momentum in the spinning black hole background, which is consistent with the results of Ref.\cite{Fujimoto:2021xix}.

The original Hawking temperature and entropy of the spinning black hole background are \cite{Hawking:1998kw}
\begin{equation}
\label{eqc2b}
\begin{split}
T=\frac{1}{2\pi} \left( r_+ (1+\frac{r^2_+}{L^2}) \left(\frac{1}{r^2_+ +a^2}+  \frac{1}{r^2_+ +b^2}  \right) -\frac{1}{r_+} \right),
\end{split}
\end{equation}
and
\begin{equation}
\label{eqc2n}
\begin{split}
S=\frac{\pi^2}{2\Xi_{a} \Xi_{b}}\frac{(r^2_+ +a^2)(r^2_+ +b^2)}{r_+},
\end{split}
\end{equation}
respectively. The horizon is defined by the largest root $r_+$ to the equation $\Delta=0$ (Eq.(\ref{eqc11})). $a$ and $b$ represent angular momentum parameters.

The free energy of black hole can be obtained by
\begin{equation}
\label{eqc2c}
\begin{split}
F(r)=\int^{r_+}_r S(x)T'(x)dx.
\end{split}
\end{equation}

\begin{figure}[H]
    \centering
      \setlength{\abovecaptionskip}{0.1cm}
    \includegraphics[width=15cm]{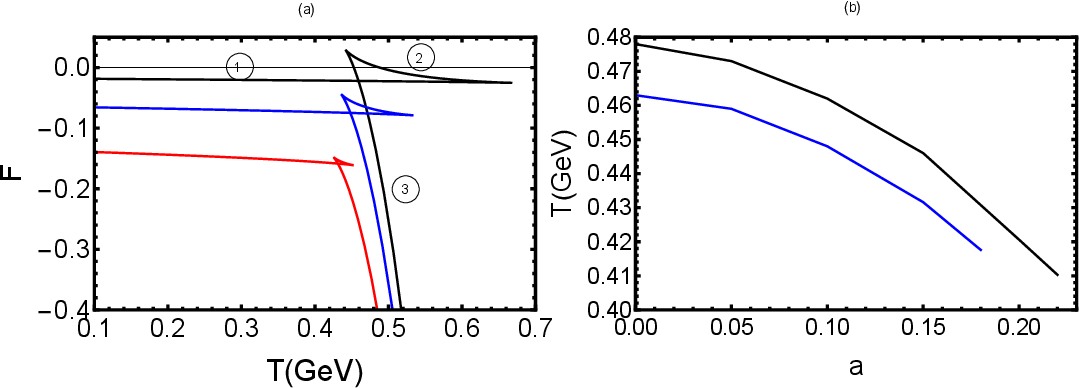}
   \caption{\label{fig7} (a) Black hole free energy $F$ versus temperature $T$ with different angular momentum parameter $a$ when $b=0.1$. The black, blue and red lines denote $a = 0.05$, $0.1$ and $0.15$, respectively. (b) First order phase transition temperature versus angular momentum parameter $a$ with different values of $b$. The black and blue lines denote $b = 0.05$ and $0.1$, respectively.}
\end{figure}

In Fig.~\ref{fig7}(a), we plot the black hole free energy $F$ versus temperature $T$. We observed that the free energy becomes multivalued when small values of $a$ and $b$ are considered. In such cases, a swallow-tailed shape appears, indicating a first-order phase transition occurs. Additionally, the small black hole (\textcircled{1}), unstable phase (\textcircled{2}), and large black hole (\textcircled{3}) coexist. The kink of the swallow-tailed shape represents a transition from small to large black holes, which is dual to the confinement-deconfinement phase transition from holography [58]. From Fig. 7(a), we can determine the temperature at which the first-order phase transition occurs. Furthermore, the swallow-tailed shape gradually disappears as increasing $a$ and vanishes at a critical point $a_c$, which indicates the phase transition temperature decreases with increasing $a$. In Fig. 7(b), we plot the first-order phase transition temperature against the angular momentum $a$ for different values of $b$. It is clear that the phase transition temperature decreases with increasing angular momentum $a$. Moreover, the blue line lies below the black line, indicating that the angular momentum $b$ suppresses the phase transition temperature. We can summarize that the first-order phase transition temperature decreases with angular momentum, which is consistent with the findings of Ref. \cite{Fujimoto:2021xix}.

\subsection*{Appendix B}

In the appendix B, we calculate analytically the quarkonium free energy for small temperature and for small angular momentum.

In transverse case, the perturbation expansion of the quarkonium free energy (Eq.(\ref{eq15})) for small angular momentum is

\begin{equation}
\label{eqc2c1}
\begin{split}
F_{Q\bar{Q}1} & =\text{\ensuremath{\frac{\sqrt{\lambda}}{\pi}}}\int_{0}^{z_{c}}\sqrt{\frac{A(z)B(z)}{A(z)-A(z_{c})}}dz,\\
& =\text{\ensuremath{\frac{\sqrt{\lambda}}{\pi}}}\int_{0}^{z_{c}}\sqrt{\frac{z_{c}^{4}(-1+\pi^{4}T^{4}z^{4})}{z^{4}(z^{4}-z_{c}^{4})}}dz+a^{2}\text{\ensuremath{\frac{\sqrt{\lambda}}{\pi}}}\int_{0}^{z_{c}}\frac{2\pi^{4}T^{4}z^{4}\sqrt{\frac{z_{c}^{4}(-1+\pi^{4}T^{4}z^{4})}{z^{4}(z^{4}-z_{c}^{4})}}}{-1+\pi^{4}T^{4}z^{4}}dz+\mathcal{O}(a^{4}),\\
& = F_1 +a^2 F'_1 +\mathcal{O}(a^{4}).
\end{split}
\end{equation}

Here, we take temperature $T=100/\pi$ and $z_c= 0.0085$. Then, we can obtain $F'_1\approx -28.91$. This indicates that the angular momentum suppresses the free energy in transverse case.

In transverse case, the perturbation expansion of the interquark distance (Eq.(\ref{eq14})) for small angular momentum is
\begin{equation}
\label{eqc2c2}
\begin{split}
L_1 & =2\int_{0}^{z_{c}}\sqrt{\frac{A(z_{c})B(z)}{A(z)^{2}-A(z)A(z_{c})}}dz,\\
 & =2\int_{0}^{z_{c}}[\sqrt{-\frac{z^{4}(-1+\pi^{4}T^{4}z_{c}^{4})}{(-1+\pi^{4}T^{4}z^{4})(z^{4}-z_{c}^{4})}}\\
 & +\frac{\sqrt{-\frac{z^{4}(-1+\pi^{4}T^{4}z_{c}^{4})}{(-1+\pi^{4}T^{4}z^{4})(z^{4}-z_{c}^{4})}}(2\pi^{4}T^{4}z^{4}-3\pi^{4}T^{4}z_{c}^{4}+\pi^{8}T^{8}z^{4}z_{c}^{4})}{2(-1+\pi^{4}T^{4}z^{4})(-1+\pi^{4}T^{4}z_{c}^{4})}a^{2}+\mathcal{O}(a^{4})]dz.
 \end{split}
\end{equation}

In transverse case, the perturbation expansion of the quarkonium free energy for small temperature is
\begin{equation}
\label{eqc2c3}
\begin{split}
F_{Q\bar{Q}2} & =\text{\ensuremath{\frac{\sqrt{\lambda}}{\pi}}}\int_{0}^{z_{c}}[\sqrt{-\frac{z_{c}^{4}}{z^{4}(z^{4}-z_{c}^{4})}}+\frac{(1+a^{2})\pi^{4}z^{4}\sqrt{-\frac{z_{c}^{4}}{z^{4}(z^{4}-z_{c}^{4})}}}{2(-1+a^{2})^{3}}T^{4}+\mathcal{O}(T^{5})]dz,\\
& = F_2 +T^4 F'_2 +\mathcal{O}(T^{5}).
 \end{split}
\end{equation}

Here, we take temperature $a=0$ and $z_c= 0.0085$. Then, we can obtain $F'_2\approx -5.70\times10^{-6}$. This indicates that the temperature suppresses the free energy in transverse case.

In transverse case, the perturbation expansion of the interquark distance for small temperature is
\begin{equation}
\label{eqc2c4}
\begin{split}
L_2=2\int_{0}^{z_{c}}[\sqrt{-\frac{z^{4}}{z^{4}-z_{c}^{4}}}+\frac{\pi^{4}\sqrt{-\frac{z^{4}}{z^{4}-z_{c}^{4}}}(-z^{4}+a^{4}z^{4}+z_{c}^{4})}{2(-1+a^{2})^{3}}T^{4}+\mathcal{O}(T^{5})]dz.
 \end{split}
\end{equation}

In parallel case, the perturbation expansion of the quarkonium free energy for small angular momentum is
\begin{equation}
\label{eqc2c5}
\begin{split}
F_{Q\bar{Q}3} &=\text{\ensuremath{\frac{\sqrt{\lambda}}{\pi}}}\int_{0}^{z_{c}}[\sqrt{\frac{z_{c}^{4}(-1+\pi^{4}T^{4}z^{4})}{z^{4}(z^{4}-z_{c}^{4})}}+\frac{3\pi^{4}T^{4}z^{4}\sqrt{\frac{z_{c}^{4}(-1+\pi^{4}T^{4}z^{4})}{z^{4}(z^{4}-z_{c}^{4})}}}{2(-1+\pi^{4}T^{4}z^{4})}a^{2}+\mathcal{O}(a^{4})]dz,\\
& = F_3 +a^2 F'_3 +\mathcal{O}(a^{4}).
 \end{split}
\end{equation}

We take temperature $T=100/\pi$ and $z_c= 0.0085$. Then, we can get $F'_3\approx -21.68$. This indicates that the angular momentum suppresses the free energy in parallel case.

In parallel case, the perturbation expansion of the interquark distance for small angular momentum is
\begin{equation}
\label{eqc2c6}
\begin{split}
L_3 & =2\int_{0}^{z_{c}}[\sqrt{-\frac{z^{4}(-1+\pi^{4}T^{4}z_{c}^{4})}{(-1+\pi^{4}T^{4}z^{4})(z^{4}-z_{c}^{4})}}\\
&+\frac{\pi^{4}T^{4}\sqrt{-\frac{z^{4}(-1+\pi^{4}T^{4}z_{c}^{4})}{(-1+\pi^{4}T^{4}z^{4})(z^{4}-z_{c}^{4})}}(z^{4}-2z_{c}^{4}+\pi^{4}T^{4}z^{4}z_{c}^{4})}{2(-1+\pi^{4}T^{4}z^{4})(-1+\pi^{4}T^{4}z_{c}^{4})}a^{2}+\mathcal{O}(a^{4})]dz.
 \end{split}
\end{equation}

In parallel case, the perturbation expansion of the quarkonium free energy for small temperature is
\begin{equation}
\label{eqc2c7}
\begin{split}
F_{Q\bar{Q}4}
 & =\text{\ensuremath{\frac{\sqrt{\lambda}}{\pi}}}\int_{0}^{z_{c}}[\sqrt{-\frac{z_{c}^{4}}{z^{4}(z^{4}-z_{c}^{4})}}+\frac{\pi^{4}z^{4}\sqrt{-\frac{z_{c}^{4}}{z^{4}(z^{4}-z_{c}^{4})}}}{2(-1+a^{2})^{3}}T^{4}+\mathcal{O}(T^{5})]dz,\\
 & = F_4 +T^4 F'_4 +\mathcal{O}(T^{5}).
 \end{split}
\end{equation}

We take temperature $a=0$ and $z_c= 0.0085$. Then, we can obtain $F'_4\approx -5.70\times10^{-6}$. This indicates that the temperature suppresses the free energy in parallel case.

In parallel case, the perturbation expansion of the interquark distance for small temperature is
\begin{equation}
\label{eqc2c8}
\begin{split}
L_4=2\int_{0}^{z_{c}}[\sqrt{-\frac{z^{4}}{z^{4}-z_{c}^{4}}}+\frac{\pi^{4}\sqrt{-\frac{z^{4}}{z^{4}-z_{c}^{4}}}(-z^{4}+2a^{4}z^{4}+z_{c}^{4}-a^{2}z_{c}^{4})}{2(-1+a^{2})^{3}}T^{4}+\mathcal{O}(T^{5})]dz.
 \end{split}
\end{equation}

One can find the interquark distance $L=L_1=L_2=L_3=L_4$ at zero temperature and zero angular momentum. Considering the transformation: $x=\frac{z}{z_{c}},\ dz=z_{c}dx,\ y=x^{4}$, one can obtain the expression of the interquark distance is

\begin{equation}
\label{eqc2c9}
\begin{split}
L& =2\int_{0}^{z_{c}}\sqrt{-\frac{z^{4}}{z^{4}-z_{c}^{4}}}dz,\\
 & =2\int_{0}^{z_{c}}\sqrt{\frac{x^{4}}{1-x^{4}}}dz,\\
 & =\frac{z_{c}}{2}\int_{0}^{1}y^{-\frac{1}{4}}(1-y)^{-\frac{1}{2}}dy,\\
 & =\frac{z_{c}}{2}B(\frac{3}{4},\frac{1}{2}),
 \end{split}
\end{equation}
where the beta function is defined by $B(a,b)=\int_{0}^{1}x^{a-1}(1-x)^{b-1}dx=\frac{\Gamma(a)\Gamma(b)}{\Gamma(a+b)}$.

The free energy $F_{Q\bar{Q}}=F_{Q\bar{Q}1}=F_{Q\bar{Q}2}=F_{Q\bar{Q}3}=F_{Q\bar{Q}4}$ at zero temperature and zero angular momentum, and the expression of free energy is
\begin{equation}
\label{eqc2c10}
\begin{split}
F_{Q\bar{Q}}& =\text{\ensuremath{\frac{\sqrt{\lambda}}{\pi}}}(\int_{0}^{z_{c}}[\sqrt{-\frac{z_{c}^{4}}{z^{4}(z^{4}-z_{c}^{4})}}-\sqrt{A(z \rightarrow0)}]dz-\int_{z_{c}}^{\infty} \sqrt{A(z \rightarrow0)}),\\
 & =\text{\ensuremath{\frac{\sqrt{\lambda}}{\pi}}}\frac{z_{c}^{-1}}{4}B(-\frac{1}{4},\frac{1}{2}),\\
 & =-\frac{1}{L}\frac{\sqrt{\lambda}(2\pi)^{2}}{\Gamma^{4}(\frac{1}{4})},
 \end{split}
\end{equation}
where the $\sqrt{A(z \rightarrow0)}$ is used to eliminate the divergence of free energy. We can find the free energy behaves as $F\propto -1/L$.

\end{document}